\documentstyle[12pt]{article}
\newcommand{\lDa}{\hbox to0pt{\raise7pt\hbox{$\leftarrow$}\hss}D}
\newcommand{\lpa}{\hbox to0pt{\raise7pt\hbox{$\leftarrow$}\hss}\partial}
\newcommand{\slD}{\hbox{$D\kern-7pt/$}}
\newcommand{\slA}{\hbox{$A\kern-7pt/$}}
\newcommand{\slDa}{\hbox to0pt{\raise7pt\hbox{$\leftarrow$}\hss}\slD}

\begin{document}
\title{Radiative corrections for the correlator of (0$^{++}$,1$^{-+}$)
light hybrid currents}
\author {{\small  H.Y. Jin and J.G. K\"orner
\thanks{jhy@thep.physik.uni-mainz.de,
koerner@thep.physik.uni-mainz.de}}\\
{\small Institut f\"ur Physik,
Johannes Gutenberg-Universit\"at,
Staudinger Weg 7, D 55099 Mainz , Germany }\\}
\date {}
\maketitle
\begin{center}
\begin{abstract}
We calculate the radiative corrections to the current-current 
correlator of the hybrid current  $g\bar
q(x)\gamma_{\nu}iG_{\mu\nu}^aT^aq(x)$. Based on this new result we use 
the QCD sum rule approach to
estimate lower bounds on the 
masses  of the $J^{PC}$=$1^{-+}$ and $0^{++}$ light hybrids.
\end{abstract}
\end{center}
\section{Introduction}
\par Mesons with exotic quantum number have attracted  a great deal of 
attention
in low energy strong interaction physics. They constitute another
intrinsic 
construction of  matter beyond the quark model  in  
QCD. Although these mesons have not been  confirmed yet, recent
experiments indeed give some evidence for their possible existence.  
 The E852 Collaboration at BNL  has reported a $J^{PC}=1^{-+}$
isovector resonance
$\hat{\rho}(1405)$ in the reaction $\pi^-p\rightarrow\eta\pi^0n$, 
with a mass of $1370\pm 16^{+50}_{-30}$ MeV and 
a width of $385\pm 40^{+65}_{-105}$ MeV \cite{bnl1}.
This state appears to have been confirmed by the Crystal Barrel
Collaboration  in $p\bar p$ annihilation  with a mass of $1400\pm20\pm20$ MeV 
and a width of $310\pm50^{+50}_{-30}$ MeV \cite{bnl11}. E852 lays claim 
to  another $J^{pc}=1^{-+}$ isovector state
$\hat{\rho}(1600)$ in the reaction 
$\pi^-p\rightarrow\pi^+\pi^-\pi^-p$,  with a mass and width of
$1593\pm8$ MeV and $168\pm20$ MeV, resp., which  decays into
$\rho\pi$ \cite{bnl2}. We will have to  wait for further confirmation of
these states.\\
 
The mass values for the exotic $1^{-+}$ state reported by the different
experiments  disagree with most 
theoretical predictions. The flux-tube model predicts 
the lowest-lying $1^{-+}$ hybrid meson to have a mass of  1.9 GeV
\cite{flux}, which is consistent with lattice QCD studies which 
predict the lightest 
exotic hybrid  $1^{-+}$ to have a mass of 2.0 GeV \cite{latt}. Besides, 
the flux-tube model also predicts that the $b_1\pi$ and $f_1\pi$ modes 
dominate in $1^{-+}$ hybrid meson decay \cite{page}, in contradiction 
with the experiments in which the mode $\rho\pi$ is dominant. 
Differing from the flux-tube model predictions calculations based on the 
 QCD
sum rule (QCDSR) approach  seem to  be closer to
experimental results \cite{shif}. The preliminary results of QCDSR show
that the lightest exotic hybrid meson has a mass around
$1.6$ GeV and a dominant  $\rho\pi$ decay mode \cite{narison}. 
This is consistent with the second $1^{-+}$ state claimed by E852. 
The discrepancies between the different model predictions 
 may come from different sources.  The flux-tube
model uses a non-relativistic linear potential model, which is not so suitable 
for the light-mass system. On the other hand, many effects, such as higher 
order terms in the OPE and radiative corrections, may affect the QCDSR
predictions.  Further theoretical studies are obviously necessary.\\
 
In this paper, we  calculate the radiative corrections to the 
current-current correlator of the  hybrid  current $\bar
q(x)\gamma_{\nu}igG_{\mu\nu}^aT^aq(x)$. 
Using this new result, we recalculate 
the  masses of the  $1^{-+}$ and $0^{++}$ hybrids via the standard QCDSR 
method.
We find that, including the radiative corrections, QCDSR  have 
less  room to accommodate  the recent experimental data. \\ 

\section{Renormalization of the current
$j_\mu=\bar q\gamma_{\nu}igG_{\mu\nu}^aT^aq$}
The operator-mixing problems associated with the renormalization of 
composite operators have been discussed sometimes ago \cite{ren}. For
instance, a given gauge invariant operator can mix with other 
gauge invariant operators, with non-gauge invariant operators which vanish 
by equations of motion and with  operators containing ghosts. The mixing
operators must have the same CP quantum number and the same dimension as
the original one. 
In our case, the complete set of operators which can mix with 
$\bar q\gamma_{\nu}igG_{\mu\nu}^aT^aq$ is given  by
\begin{eqnarray}\label{cur}\nonumber
j^1_\mu&=&\bar q\gamma_{\nu}igG_{\mu\nu}q,\\
\nonumber
j^2_\mu&=&\bar q (\vec{D}_\mu \vec{\slD}-\slDa\lDa_\mu) q,\\
\nonumber
j^3_\mu&=&\bar q (\gamma_{\mu}\sigma_{\alpha\beta}gG_{\alpha\beta}
-gG_{\alpha\beta}\sigma_{\alpha\beta}\gamma_\mu)q,\\
j^4_\mu&=&\bar q (\gamma_{\mu}\vec{\slD}\vec{\slD}-
{\slDa}{\slDa}\gamma_{\mu})q,\\
\nonumber
j^5_\mu&=&\bar q(\gamma_{\mu}ig\slA\vec{\slD}+\slDa
ig\slA\gamma_{\mu})q,\\
\nonumber
j^6_\mu&=&\bar q (igA_\mu\vec{\slD}+\slDa igA_\mu)q.
\end{eqnarray}
Note that there is no dimension-five operator containing ghost fields 
in the set (\ref{cur}). In (\ref{cur}) we have defined 
$\sigma_{\alpha\beta}=\frac{i}{2}[\gamma_\alpha,\gamma_\beta]$ and 
the covariant derivatives 
$\vec{D}_\mu=\vec\partial_\mu+igA_\mu$, $\lDa_\mu=\lpa_\mu-igA_\mu$,
which  act on right and left fields, respectively.    
The fields and couplings in (\ref{cur}) are bare.  Only  
$j^1_\mu$ and $j^3_\mu$ are physical currents. The other currents 
 correspond to  so-called 
nuisance operators  which vanish by the equations of motion. 

Renormalizing the composite operator corresponding to $j^1_\mu$ we 
obtain 
\begin{equation}\label{court}
[j^1_\mu]=Z_1j^1_\mu
+Z_2j^2_\mu+Z_3j^3_\mu+Z_4j^4_\mu+Z_5j^5_\mu+Z_6j^6_\mu,
\end{equation}
In order to determine 
the coefficients $Z_i(i=1,6)$, we insert the currents into 1PI diagrams 
and extract the ultraviolet divergences. We find that the divergences
associated 
with the $\bar q qg$-vertex are sufficient to determine all counterterms.   
The relevant diagrams are shown in Fig.1.  In Feynman gauge, we obtain 
\begin{eqnarray}\nonumber
Z_1&=&\displaystyle{1+\frac{113}{18}\frac{g^2}{16\pi^2}\frac{1}{\epsilon}},\\
\nonumber
Z_2&=&\displaystyle{-\frac{4}{9}\frac{g^2}{16\pi^2}\frac{1}{\epsilon}},\\
\nonumber
Z_3&=&\displaystyle{-\frac{49}{72}\frac{g^2}{16\pi^2}\frac{1}{\epsilon}},\\
Z_4&=&\displaystyle{-\frac{8}{9}\frac{g^2}{16\pi^2}}\frac{1}{\epsilon},\\
\nonumber
Z_5&=&\displaystyle{\frac{19}{72}\frac{g^2}{16\pi^2}\frac{1}{\epsilon}},\\
\nonumber
Z_6&=&\displaystyle{\frac{35}{36}\frac{g^2}{16\pi^2}\frac{1}{\epsilon}},
\end{eqnarray}
where we use dimensional regularization. 

\section{Next-to leading order correction to the current-current
correlator}
 Let us consider the current-current correlator 
\begin{eqnarray}\label{cor}
\Pi_{\mu\nu}(q^2)&=&i\int d^4x e^{iqx} \langle 0|
T\{j_{\mu}(x),j^+_{\nu}(0)\}|0 \rangle \\
&=&\displaystyle{(\frac{q_\mu 
q_\nu}{q^2}-g_{\mu\nu})\Pi_v(Q^2)+\frac{q_\mu
q_\nu}{q^2}\Pi_s(q^2)}\nonumber
\end{eqnarray}
where $j_{\mu}(x)=\bar u(x)\gamma_{\nu}igG_{\mu\nu}^aT^au(x)$. The
invariants
$\Pi_v(Q^2)$ and $\Pi_s(Q^2)$ correspond to the contributions from
$1^{-+}$ and $0^{++}$ states and their excited states, respectively. The
leading
order  
contribution to (\ref{cor})
including the quark and gluon condensate contributions has already been
given in \cite{lead}
\begin{eqnarray}\label{lo}
\Pi^0_v(q^2)& = &\displaystyle{-\left[\frac{\alpha_s}{240\pi^3}(q^2)^3+
\frac{1}{36\pi}q^2(\langle\alpha_sG^2\rangle+8\alpha_s\langle m\bar uu)
\rangle
\right]\ln(\frac{-q^2}{u^2})}\\\nonumber
&&\displaystyle{
+\left[\frac{4\pi}{9}\alpha_s\langle\bar uu\rangle^2
+\frac{1}{192\pi^2}g^3\langle G^3 \rangle-\frac{83\alpha_s}{1728\pi}m
\langle\bar uGu\rangle\right ], }\\\nonumber
\Pi^0_s(q^2)&=&\displaystyle{
\left[-\frac{\alpha_s}{480\pi^3}(q^2)^3+(\frac{\alpha_s}{3\pi}\langle 
m\bar uu \rangle+\frac{\langle\alpha_sG^2\rangle}{24\pi})q^2+
\frac{m^2}{8\pi}\langle\alpha_s G^2 \rangle+
\frac{11\alpha_s}{72\pi}m\langle\bar uGu\rangle\right 
]\ln(\frac{-q^2}{u^2})}\\\nonumber
& &\displaystyle{
+8\pi\alpha_s\langle\bar uu\rangle^2.}
\end{eqnarray}

There is no difference for isovector and isoscalar currents in this order. 
The next-to-leading order correction to the perturbative 
part of $\Pi_v(q^2)$ and $\Pi_s(q^2)$  can be obtained by calculating
the Feynman diagrams in Fig.2, where Figs.2m and 2n only
contribute to the isoscalar states. The technique of the calculation 
which we use here  was firstly proposed in \cite{chet}. 

Let us now briefly comment on the radiative corrections to the
correlator of the current $g\bar q\gamma_5\gamma_{\nu}iG_{\mu\nu}^aT^aq$
with $0^{--}$ and $1^{+-}$ quantum numbers.  The isovector current 
correlator  has
the same radiative correction as the isovector  current correlator of 
the ($0^{++}$, $1^{-+}$) current.
However, the isoscalar current correlator has different radiative
corrections since diagrams Figs.2m and 
2n, which correspond to the mixing with pure gluonic states, now give zero
contributions.

Each of the diagrams  Figs.2i-2n is gauge-parameter independent by itself 
because the current is antisymmetric in the Lorentz indices. 
We have checked on gauge invariance for
the sum of diagrams Figs.2a-2h  by doing 
the calculation in a general covariant gauge and found that the result is
gauge-parameter independent. Explicit Feynman gauge results 
for the  diagrams are listed in the appendix.

In the present application only the isovector currents are of interest. 
In the $\rm\overline{MS}$-scheme,  the next-to leading corrections  
to the correlator of 
isovector currents  is given by 
\begin{eqnarray}\label{nlo}
\Pi^{1a}_v(q^2)&=& \displaystyle{-\frac{\alpha_s(\mu)}{240\pi^3}(q^2)^3
\ln(\frac{-q^2}{\mu^2})\left[(\frac{53}{4}-\frac{76}{45}n_f)
\frac{\alpha_s(\mu)}{\pi}-(\frac{35}{24}-\frac{1}{4}n_f)
\frac{\alpha_s(\mu)}{\pi}
\ln(\frac{-q^2}{\mu^2})\right]}\\
\Pi^{1a}_s(q^2)&=& \displaystyle{-\frac{\alpha_s(\mu)}{480\pi^3}(q^2)^3
\ln(\frac{-q^2}{\mu^2})\left[(\frac{2017}{216}-\frac{229}{180}n_f)
\frac{\alpha_s(\mu)}{\pi}-(\frac{35}{24}-\frac{1}{4}n_f)
\frac{\alpha_s(\mu)}{\pi}
\ln(\frac{-q^2}{\mu^2})\right]}\nonumber
\end{eqnarray} 
where we take  
the light quark mass to be zero for convenience. Although 
(\ref{lo}) include condensates which are proportional to the light quark 
mass, their contributions are very small compared to those of  
operators with the same dimension. Throughout this
calculation taking zero quark mass is a 
good approximation. 

Eq.(\ref{nlo}) does not contain the  complete next-to leading
order correction. one  also has to include the contribution from the
renormalization of the current. 
By inserting the renormalized currents  
(\ref{court}) into the correlator
\begin{equation}\label{1b}
i\int d^4x e^{iqx} 2Z^{\alpha_s}_i
\langle 0|T\{j^1 _{\mu}(x),j^{i+}_{\nu}(0)\}|0 \rangle,
\end{equation}
we obtain   
\begin{eqnarray}\label{nlob}
\Pi^{1b}_v(q^2)&=& \displaystyle{-\frac{\alpha_s(\mu)}{240\pi^3}(q^2)^3
\ln(\frac{-q^2}{\mu^2})((-\frac{91}{16}+\frac{39}{40}n_f)
\frac{\alpha_s(\mu)}{\pi}+(\frac{35}{36}-\frac{1}{6}n_f)
\frac{\alpha_s(\mu)}{\pi}
\ln(\frac{-q^2}{\mu^2}))}\\
\Pi^{1b}_s(q^2)&=& \displaystyle{-\frac{\alpha_s(\mu)}{480\pi^3}(q^2)^3
\ln(\frac{-q^2}{\mu^2})((-\frac{679}{144}+\frac{97}{120}n_f)
\frac{\alpha_s(\mu)}{\pi}+(\frac{35}{36}-\frac{1}{6}n_f)
\frac{\alpha_s(\mu)}{\pi}
\ln(\frac{-q^2}{\mu^2}))}.\nonumber
\end{eqnarray} 
In Eq.(\ref{1b}) we have introduced the notation $Z^{\alpha_s}_1=
Z_1Z_g-1$
and $Z^{\alpha_s}_i=Z_i(i=2,..6)$.
For the original hybrid current $j^1_\mu$ one also has to include the
counterterm $Z_g$ resulting from the renormalization of the bare coupling 
constant $g$.  
The final result for the radiative corrections to the correlator is
finally obtained by adding the contributions of (\ref{nlo}) and
(\ref{nlob}).

\section{Sum rules  for  $1^{-+}$ and $0^{++}$ hybrid mesons}
For sum rule applications we need the spectral density associated with the 
current-current correlator.  
The spectral density $\rho_v(s)=Im\Pi_v(s)$ is defined via  the standard
dispersion relation 
\begin{equation}\label{dr}
\Pi_v(q^2)=\displaystyle{\frac{(q^2)^n}{\pi}
\int^\infty_0ds\frac{\rho_v(s)}{s^n(s-q^2)}+\sum^{n-1}_{k=0}a_k(q^2)^k},
\end{equation}
where the $a_k$ are appropriate subtraction constants to render
Eq.(\ref{dr}) finite.

From (\ref{lo}), (\ref{nlo}) and (\ref{nlob}) we obtain
\begin{eqnarray}\label{QCD}
\rho_v(s)&=& \displaystyle{\frac{\alpha_s(\mu)}{240\pi^2}s^3
(1+\frac{1301}{240}\frac{\alpha_s(\mu)}{\pi}-\frac{17}{36}\frac{\alpha_s(\mu)}{\pi}
\ln(\frac{s}{\mu^2}))+
\frac{1}{36}s(\langle\alpha_sG^2\rangle+8\alpha_s\langle m\bar uu
\rangle)
}\\
\rho_s(s)&=&\displaystyle{
\frac{\alpha_s(\mu)}{480\pi^2}s^3
(1+\frac{6979}{2160}
\frac{\alpha_s(\mu)}{\pi}-\frac{17}{36}\frac{\alpha_s(\mu)}{\pi}
\ln(\frac{s}{\mu^2}))}\\\nonumber
&&\displaystyle{
-(\frac{\alpha_s}{3}\langle
m\bar uu \rangle+\frac{\langle\alpha_sG^2\rangle}{24})s-
\frac{m^2}{8}\langle\alpha_s G^2 \rangle-
\frac{11\alpha_s}{72}m\langle\bar uGu\rangle },\nonumber
\end{eqnarray}
where we have set $n_f=3$.

On the other hand, the spectral density is saturated by 
narrow physical resonances and the continuum. We therefore write 
\begin{equation}\label{phys}
\displaystyle{(\frac{q_\mu
q_\nu}{q^2}-g_{\mu\nu})}\rho_v(s)=\displaystyle{\sum_R
\langle 0| j_\mu|R\rangle\langle R|j_\nu |0\rangle}
\pi\delta(s-m^2_R)+continuum,
\end{equation}
where we have assumed that the mass $m_R$  is much larger than the 
decay width of the hybrid, so that the imaginary part of the propagator 
has been replaced by $\pi\delta(s-m^2_R)$. 

In order to extract information on the lowest-lying resonance,
it is usually  assumed that the lowest-lying resonance dominates  
the spectral density. The contribution of  higher excited states 
can be suppressed by applying  the Borel transformation $\hat L_M$ 
to  both sides of Eq. (\ref{dr}). One thus has
\begin{equation}\label{bor}  
R_0(\tau)=\displaystyle{M\hat L_M 
\Pi_v(q^2)=\frac{1}{\pi}\int^\infty_0
e^{-s/M} \rho_v(s)ds}.
\end{equation}
The upper limit of the intergral can be replaced 
by a finite number $s_0$. The contributions beyond the threshold $s_0$ 
are considered to result from the continuum. $R_0(\tau)$ is 
the zeroth moment. Higher order moments are defined by 
$R_k=(M^2\frac{\partial}{\partial M})^k R_0(M)$. Resonance masses  can be
obtained by taking the ratio $m_R^2= \frac{R_{k+1}}{R_k}$ with the
assumption that
only a single narrow resonance dominates.  In principle 
any value  of $k$ can be chosen to determine the resonance masses.  
However, since we have to truncate the series of the power expansion,  
using higher moments 
will damage the convergence of the OPE. Besides, it is also
arbitrary to define the scalar function  
$\Pi^k_v(q^2)=\frac{1}{(q^2)^k}\Pi_v(q^2)$ 
by extracting 
a general tensor factor
\begin{equation}\label{fact} 
\displaystyle{(q^2)^k(\frac{q_\mu q_\nu}{q^2}-g_{\mu\nu})}
\end{equation}
from the correlator
$(\ref{cor})$. Although (\ref{bor}) can be considered as a higher moment 
of $\Pi^k_v(q^2)$, to the order of OPE that we are  considering,
the sum rules for $\Pi^k_v(q^2)$ and 
$\Pi_v(q^2)$  are obviously different. For instance, as pointed out
in \cite{lead}, the dimension-six operators of (\ref{lo}) do not
contribute to sum rule (\ref{bor}), while they play an important role in 
stabilizing the sum rule of  $\Pi^k_v(q^2)$ in the case of $k=1$.  
We shall consider the two cases $k=0$ and $k=1$ in turn.

The single particle matrix elements contributing to (\ref{phys})  are
parametrized as 
\begin{eqnarray}\label{mat}
\langle 0|j_\mu|V\rangle&=&i\epsilon_\mu f_vm_v^3.\\
\langle 0|j_\mu|S\rangle&=&ip_\mu f_sm_s^2.\nonumber
\end{eqnarray}
In the narrow resonance approximation the sum rules are  independent of
the matrix 
element (\ref{mat}). When the decay width of the resonance is comparable 
with its mass, the approximation  (\ref{phys}) is no longer valid. Then 
the information contained in (\ref{mat}) may become important. We will
comment on this 
later on.  By using (\ref{QCD})-(\ref{mat}) we obtain 
\begin{equation}\label{mass}
m_{v,s}^2=\displaystyle{\frac{\int_0^{s_0} e^{-s\tau}s\rho_{v,s}(s)ds}
{\int_0^{s_0}e^{-s\tau}\rho_{v,s}(s)ds}}
\end{equation}
for the sum rule  (\ref{bor}) and 
\begin{eqnarray}\label{mass1}
m_v^2&=&\displaystyle{\frac{\int_0^{s_0} e^{-s\tau}\rho_v(s)ds}
{\int_0^{s_0}e^{-s\tau}\rho_v(s)ds/s-\frac{4\pi^2}{9}\alpha_s\langle\bar
uu\rangle^2
-\frac{1}{192\pi}g^3\langle G^2 \rangle+\frac{83}{1728}\alpha_sm
\langle\bar uGu\rangle}} \\
m_s^2&=&\displaystyle{\frac{\int_0^{s_0} e^{-s\tau}\rho_s(s)ds}
{\int_0^{s_0}e^{-s\tau}\rho_s(s)ds/s-8\pi^2\alpha_s\langle\bar
uu\rangle^2}}\nonumber
\end{eqnarray}
for the sum rule for $\Pi^k_v(q^2)$ with $k=1$ , where the
spectral
densities $\rho_{v,s}(s)$ are given
in
(\ref{QCD}).
The  various parameters entering in (\ref{mass})-(\ref{mass1}) 
are specified as \cite{par}
\begin{eqnarray}
\Lambda_{QCD} =0.25GeV,&m = 0.01GeV,&m\langle\bar uu\rangle
=-\frac{1}{4}f_\pi^2m_\pi^2\\\nonumber
\langle\alpha_sG^2\rangle=0.04Gev^4,&g^3\langle G^3\rangle =
1.1GeV^2\langle\alpha_sG^2\rangle,&f_\pi=0.132GeV\\
\alpha_s(\mu)=\displaystyle{\frac{4\pi}{9\ln(\frac{\mu^2}{\Lambda^2_{QCD}})}},
&\mu=2GeV,&g\langle\bar uGu\rangle=1.5Gev^2\langle\bar uu\rangle\nonumber
\end{eqnarray}

In  Fig.3 and Fig.4 we show a mass  plot of the $1^{-+}$ state in 
its dependence on the Borel parameter $M$ in the two sum rule ratios. 
The second sum rule gives a smaller mass which we will  
take as the lower bound. 

The sensitivity of the mass to the choice of the threshold value $s_0$ is
obvious. 
The mass will go to infinity when both 
$s_0$ and $M$ go to infinity, because, when $M$ goes 
to infinity, the  Borel measure (\ref{bor}) no longer suppresses the
continuum. In order to give a reasonable estimate, we set  $s_0$ 
around $4$ GeV$^2$.   Fig.5 shows that  $m_v$ takes values around $3$ GeV 
when $s_0$ is set to infinity. In  Fig.6 one cannot find any 
stable point in the two-dimensional ($s_0$,$M$) space.  
Therefore, before fixing $s_0$, we 
cannot make any precise prediction for the hybrid mass. However, 
if one believes that the $1^{-+}$ hybrid mass lies around 2 $GeV$, 
$s_0$ should be larger than 4 $GeV^2$. This results in a lower bound for
the $1^{-+}$ mass of 1.55 GeV(see Fig.2). 
The radiative corrections enhance 
the lower bound which gives QCDSR less room 
to accommodate recent experimental data.

The prediction of the mass is also sensitive to the form of the spectral 
density. This is mostly due to a truncated OPE.
The contributions from higher dimension operators  are very likely  
not small. The
uncertainty from  the narrow resonance approximation in (\ref{phys}) 
does not seem  to reduce this discrepancy. 
This can be checked by replacing the narrow resonance form 
$\pi\delta(s-m_R^2)$ in (\ref{phys}) by the Breit-Wigner form    
\begin{equation}
\displaystyle{\frac{\Gamma_Rm_R}{(s-m_R^2)^2+\Gamma^2m_R^2}}.
\end{equation}
$\Gamma_R$ is the width of the $1^{-+}$ hybrid. When we choose
$\Gamma_R=200MeV$ and a parametrization of the matrix element 
(\ref{mat}) in the form $\langle 0|j_\mu|V\rangle=i\epsilon_\mu 
m_R\bar f_v s^x$ with $\bar f_v$ constant and $x=0.5\div 1.5$, 
 the hybrid mass is not sensitive to using the full  Breit-Wigner form 
propagator. We show the change in the Figs.3 and 4 by 
dotted lines. The discrepancy between  the two cases k=0 and k=1 is still
big and
the mass predictions  become  somewhat larger. 

The sum rule for the $0^{++}$ hybrid is shown in Figs.7 and 8 for the two 
cases, k=0 and k=1  resp., where we set $s_0=7$ GeV$^2$.
The radiative corrections reduce the discrepancy between the two cases.  
Similar to the $1^{-+}$ case, when $s_0$ goes to infinity, the prediction
of mass is around $3$ GeV. It means that the contribution of the continuum 
dominates over that of the resonance´s. Therefore, the value
chosen for the threshold $s_0$ is
important.    

\section{Summary}
In  summary, we have calculated the next-to-leading order corrections to 
the two point  correlator of the current 
$g\bar q\gamma_{\nu}iG_{\mu\nu}^aT^aq(x)$. We recalculated  
the masses of the $1^{-+}$ and $0^{++}$ hybrids. We find that the
radiative corrections reduce the lower bound of the $1^{-+}$ mass and
leave less room for QCD sum rules to fit the recent experimental data.\\     

{\bf Note added in proof}: While preparing this paper for publication, we
became aware of a recent paper by K. Chetyrkin and S. Narison which
addresses similar problems\cite{che}. 

\vspace{1.0cm}
{\bf Acknowledgment}
We would like to thank  S. Groote, A. A. Pivovarov  and K. Chekyrkin for
very useful discussions. We would also like to thank K. Chekyrkin for 
providing us with intermediate results of the calculation\cite{che}.
The work of H.Y. J. is supported  by the  Alexander von Humboldt
foundation.

\newpage

{\bf Appendix}
In this appendix we list the results of calculating the diagrams Fig.2 in 
the Feynman gauge for the correlator (\ref{cor}).
\begin{eqnarray}\nonumber
Fig.2a&:&C\displaystyle{
\left[(\frac{637}{3840}
-\frac{3}{128}
\ln(\frac{-q^2}{\mu^2}))g_{\mu\nu}+(-\frac{143}{640}
+\frac{9}{256}\ln(\frac{-q^2}{\mu^2}))\frac{q_\mu q_\nu}{q^2}
\right ]
}\\\nonumber
Fig.2b&:& \displaystyle{C
\left [(-\frac{673}{172800}
+\frac{1}{1920}
\ln(\frac{-q^2}{\mu^2}))g_{\mu\nu}+(\frac{9}{1600}
-\frac{1}{1280}\ln(\frac{-q^2}{\mu^2}))\frac{q_\mu q_\nu}{q^2}
\right ]
}\\\nonumber
Fig.2c&:& \displaystyle{C\left [\frac{1}{480}g_{\mu\nu}+(-\frac{307}{3840}
+\frac{3}{256}\ln(\frac{-q^2}{\mu^2}))\frac{q_\mu q_\nu}{q^2}
\right ]
}\\\nonumber
Fig.2d&:& \displaystyle{C\left [(-\frac{157}{1600}
+\frac{9}{640} 
\ln(\frac{-q^2}{\mu^2}))g_{\mu\nu}+(\frac{867}{6400}
-\frac{27}{1280}\ln(\frac{-q^2}{\mu^2}))\frac{q_\mu q_\nu}{q^2}
\right ]
}\\\nonumber
Fig.2e&:& \displaystyle{C\left [(-\frac{583}{21600}
+\frac{1}{240}
\ln(\frac{-q^2}{\mu^2}))g_{\mu\nu}+(\frac{31}{800}
-\frac{1}{160}\ln(\frac{-q^2}{\mu^2}))\frac{q_\mu q_\nu}{q^2}
\right ]
}\\\nonumber
Fig.2f&:& \displaystyle{C\left [\frac{1}{640}g_{\mu\nu}+
\frac{1}{640}\frac{q_\mu q_\nu}{q^2}
\right ]
}\\\nonumber
Fig.2(g+h+i)&:& \displaystyle{C\left [(\frac{79}{1440}
-\frac{19}{2700}n_f-(\frac{1}{128}-\frac{1}{960}n_f)
\ln(\frac{-q^2}{\mu^2}))g_{\mu\nu}\right . }  \\\nonumber
&&\displaystyle{\left .+(-\frac{97}{1280}
+\frac{31}{3200}n_f
+(\frac{3}{256}-\frac{1}{640}n_f)
\ln(\frac{-q^2}{\mu^2}))\frac{q_\mu q_\nu}{q^2}
\right ]
}\\\nonumber
Fig.2j&:& \displaystyle{C\left [(-\frac{1}{24}
+\frac{1}{144}
\ln(\frac{-q^2}{\mu^2}))g_{\mu\nu}+(\frac{25}{216}
-\frac{1}{48}\ln(\frac{-q^2}{\mu^2}))\frac{q_\mu q_\nu}{q^2}
\right ]
}\\\nonumber
Fig.2k&:& \displaystyle{C\left
[-\frac{1}{720}g_{\mu\nu}+\frac{13}{3240}\frac{q_\mu q_\nu}{q^2}
\right ]
}\\\nonumber
Fig.2l&:& \displaystyle{C\left [(\frac{83}{28800}
-\frac{1}{1920}
\ln(\frac{-q^2}{\mu^2}))g_{\mu\nu}+(\frac{289}{86400}
-\frac{1}{1920}\ln(\frac{-q^2}{\mu^2}))\frac{q_\mu q_\nu}{q^2}  
\right ]
}\\\nonumber
Fig.2m&:& \displaystyle{C\left [-\frac{1}{8640}g_{\mu\nu}
-\frac{1}{1440}\frac{q_\mu q_\nu}{q^2}
\right ]
}\\\nonumber
Fig.2n&:& \displaystyle{C\left [(\frac{71}{5400}
-\frac{1}{480}
\ln(\frac{-q^2}{\mu^2}))g_{\mu\nu}+(-\frac{401}{21600}
+\frac{1}{320}\ln(\frac{-q^2}{\mu^2}))\frac{q_\mu q_\nu}{q^2}
\right ]
}\\\nonumber
\end{eqnarray}

We use the abbreviation
$C=\frac{\alpha_s(\mu)^2}{\pi^4}(q^2)^3\ln(\frac{-q^2}{\mu^2})$.

\newpage
\par
{\huge\bf Figure captions}\\
\par
Fig. 1  Feynman diagrams for the renormalization of the hybrid current.
Dots stand for the current vertices.\\
\par
\par
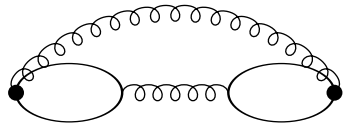
Fig. 2. Feynman diagrams for the next-to-leading calculation. Dots 
stand for the current vertices.\\
\par
\par
Fig. 3.  $1^{-+}$ hybrid mass $m_v$ versus Borel variable $M$ for the
first
sum rule
Eq.(\ref{mass}). 
Dashed line gives the result for the leading order calculation. 
Solid line includes  radiative corrections.  Dotted line gives the results
 when using 
the Breit-Wigner resonance propagator.\\
\par
\par
Fig. 4.  $1^{-+}$ hybrid mass $m_v$ versus Borel variable $M$ for 
the second sum rule Eq.(\ref{mass1})\\
\par
\par
Fig. 5.  $1^{-+}$ hybrid mass $m_v$ versus Borel variable $M$ 
 for the second sum rule Eq.(\ref{mass1})
 when $s_0$ goes to infinity.\\
\par
\par
Fig. 6.  Three-dimensional  figure of $1^{-+}$ hybrid mass $m_v$ vs. the
Borel variable $M$ and the threshold parameter $s_0$.\\
\par
\par
Fig. 7.  $0^{++}$ hybrid mass $m_s$ versus Borel variable $M$ for the
first
sum rule Eq.(\ref{mass}). Solid line inculdes radiative corrections.\\
\par
\par
Fig. 8.   $0^{++}$ hybrid mass $m_s$ versus Borel variable $M$ for the
second sum rule Eq.(\ref{mass1}). Solid line includes radiative
correction.\\
\par
\par

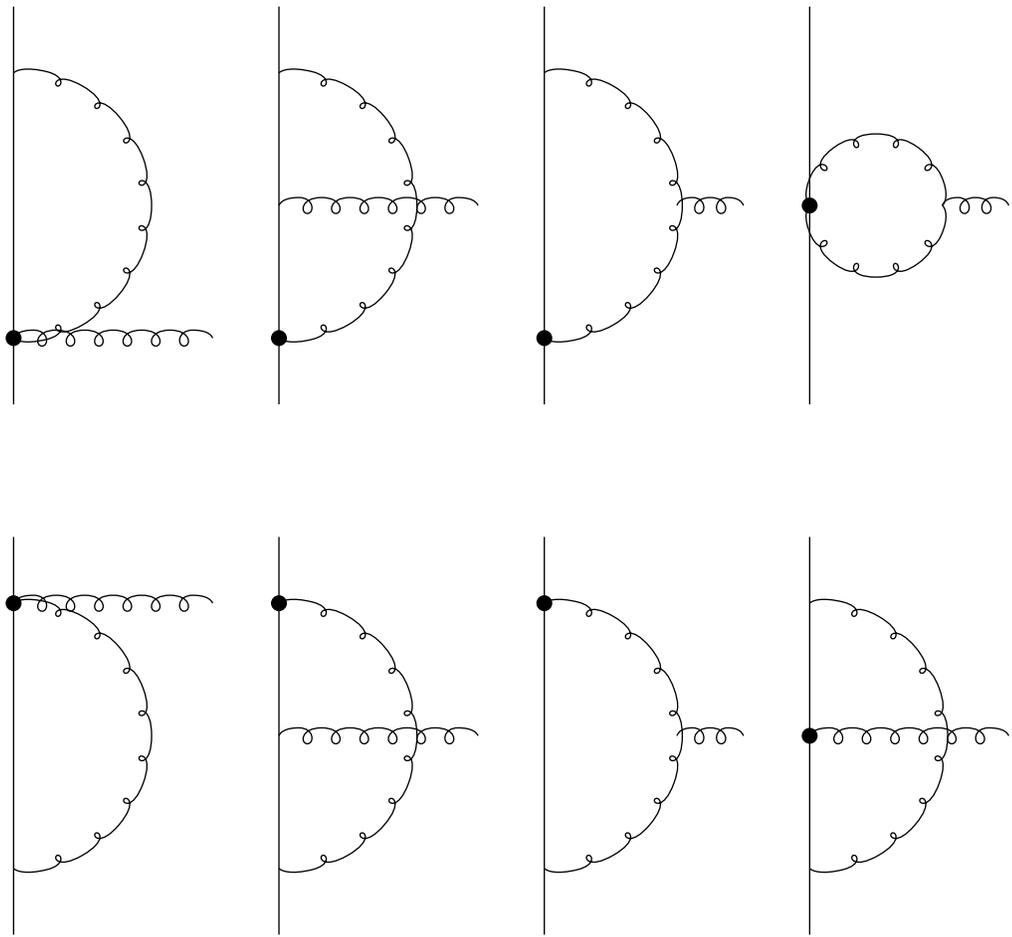
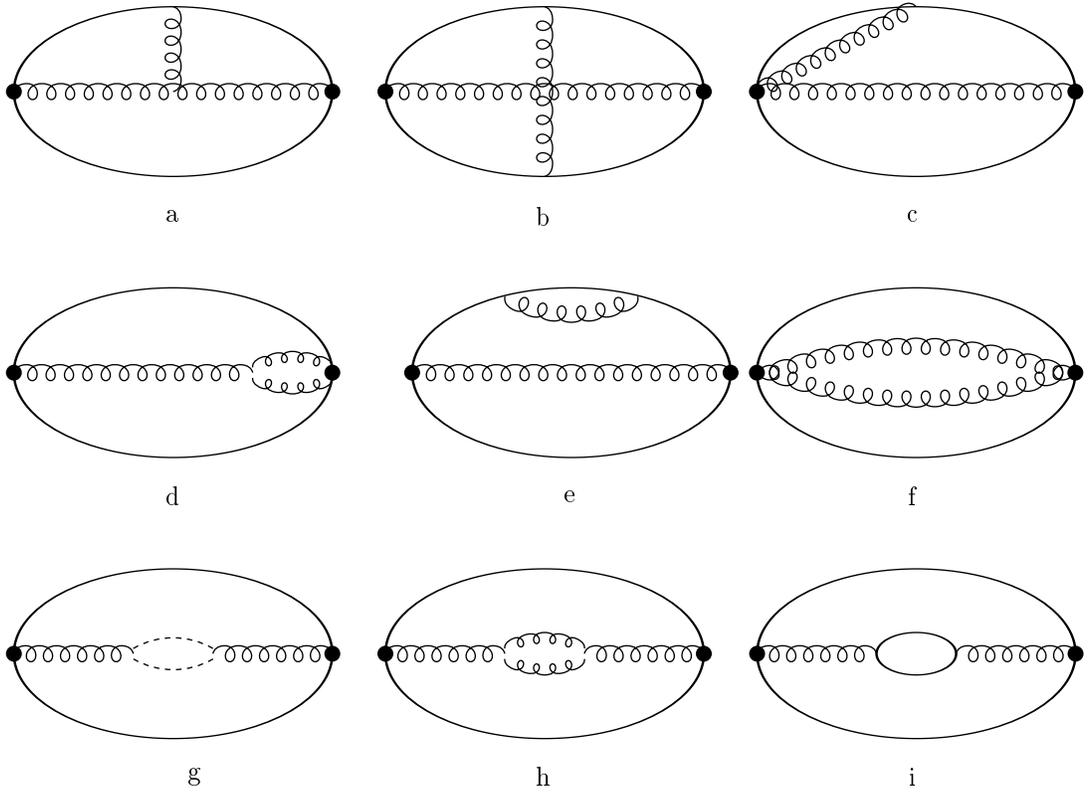
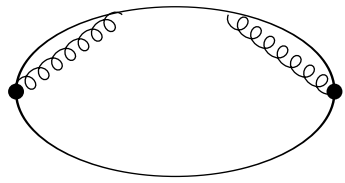
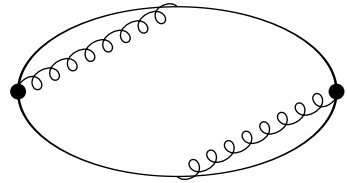
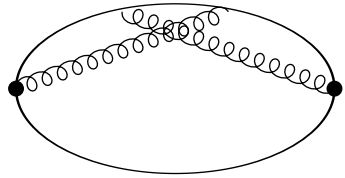
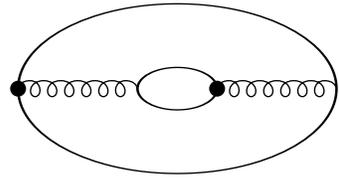
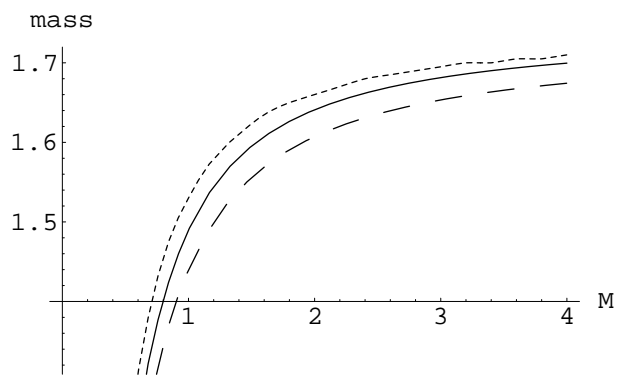
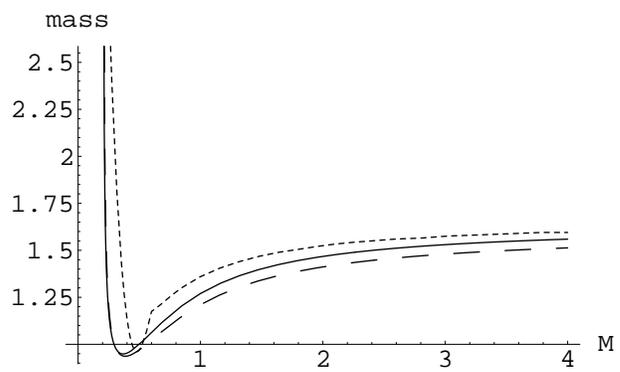
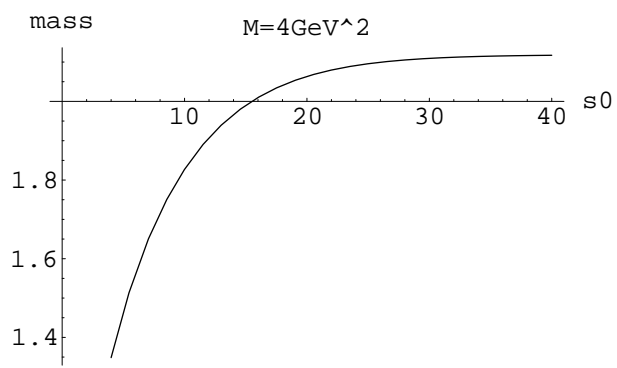
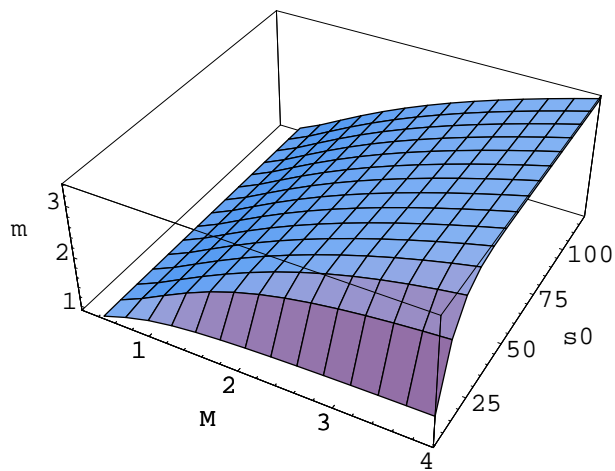
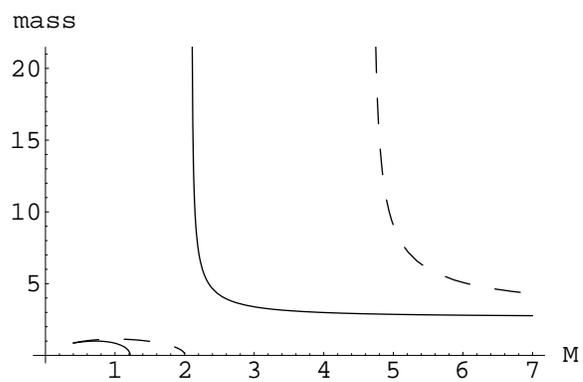
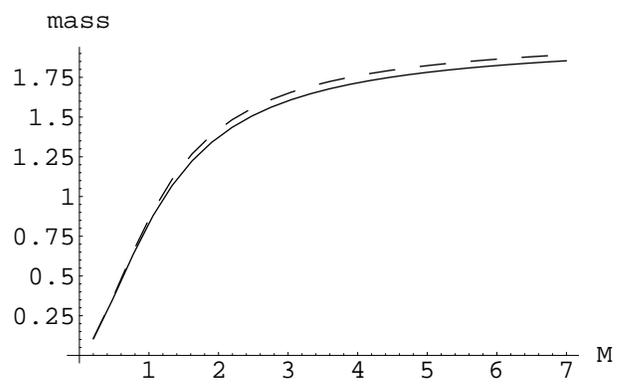
\end{document}